# Symmetry Theory of the Flexomagnetoelectric Effect in the Magnetic Domain Walls

**B. M. Tanygin**[a]

[a] Kyiv Taras Shevchenko National University, Radiophysics Faculty, Glushkov av.2, build.5, Kyiv, Ukraine, MSP 01601

*Corresponding author:* B.M. Tanygin, 64 Vladimirskaya str., Taras Shevchenko Kyiv National University, Radiophysics Faculty. MSP 01601, Kyiv, Ukraine.

*E-mail*: b.m.tanygin@gmail.com

*Phone*: +380-68-394-05-52

**Abstract.**

A local flexomagnetoelectric *(A.P. Pyatakov, A.K. Zvezdin, 2009)* effect in the magnetic domain walls (DWs) of the cubic hexoctahedral crystal has been investigated on the basis of a symmetry analysis. The strong connection between magnetic symmetry of the DW and the type of the distribution of the electric polarization was shown. Results were systemized in the scope of the DW chirality. It was shown, that new type of the local flexomagnetoelectric coupling can be experimentally detected using the external homogeneous electric field.

## 1. Introduction

Coupling mechanism between magnetic and electric subsystem in magnetoelectric materials [1] is of considerable interest to fundamentals of condensed matter physics and for the applications in the novel multifunctional devices [2]. The electric polarization can be induced by the homogeneous [3-7] and inhomogeneous [8-12] magnetization distributions. The last are micromagnetic structures like domain walls (DWs) [8,12-17] and magnetic vortexes [18]. It was shown [17] that the magnetoelectric coupling in the magnetic DWs can be described in the same manner as the ferroelectricity in the spiral magnets. Such type of the magnetoelectric interaction is described by the Lifshitz invariant-like coupling term $P_i M_j \nabla_k M_n$ [5,8,10-12]. It was called as flexomagnetoelectric interaction [19-25]. In general case such free energy term is allowed by any crystal symmetry [26]. Consequently, the electric polarization induced by the micromagnetic structure can appear in any magnetic material even in centrosymmetric one.

It is well known that magnetoelectric effects are closely related to the magnetic symmetry. This principle has been applied for the DWs with specific symmetry [8,12]. It was not compared with phenomenological description of the flexomagnetoelectric effect. The purpose of this work is to extend the symmetry classification to all possible magnetic point groups and compare results with the phenomenological investigations of the cubic $m\bar{3}m$ crystal.

## 2. Group-theoretical description of the flexomagnetoelectric coupling in the domain walls

### 2.1. Symmetry classification of the domain walls

Since DWs can be considered as thin layers, their symmetry is described by one of the 528 magnetic layer groups [28,29]. To determine the layer's physical properties continuum approximation is used which leads to point-like layer groups [30]. If continuous translation operation is considering as identity then these groups transform to magnetic point groups. It was shown [31] that there are 125 of such groups. It was found that if magnetic point group is pyroelectric and/or pyromagnetic then DW carries **P** and/or **M** respectively [32]. These criteria were derived from the conditions of the appearing of the uniform **P** [33,34] and/or **M** [35,36]. After their application to any inhomogeneous region they predict existing of even parts in functions of distribution of **P** and/or **M** [37-40]. Identification of the remained odd parts of these functions, were formulated [37-40] based on symmetry transformations which interrelate domains.

Let us choose the coordinate system $XYZ$ connected with the DW plane (axis Z is the DW plane normal direction). The distribution of the polarization is $\mathbf{P}(\mathrm{z}) = \mathbf{e_x}P_x(z) + \mathbf{e_y}P_y(z) + \mathbf{e_z}P_z(z)$. The magnetic point group of the DW contains two types of symmetry transformations. First type transformations $\mathrm{g}^{(1)}$ do not change the spatial coordinate $z$. If transformation $\mathrm{g}^{(1)}$ is a rotation around $n$-fold ($n > 1$) symmetry axis $n_z$ then it allows only $P_x(z) = P_y(z) = 0$. Second type transformation $\mathrm{g}^{(2)}$ has opposite property: $\mathrm{g}^{(2)}z = -z$. This transformation allows only even (S) or odd (A) functions $P_i(z)$, where $i = x, y, z$. If the magnetic point group of the DW is non-polar then $\mathbf{P}(z) = -\mathbf{P}(-z)$ and at least component $P_z(z)$ is non-zero. Consequently, all 125 magnetic point groups of DWs allow **P**.

If magnetic point group of DW is non-pyromagnetic, then $\mathbf{M}(\mathrm{z}) = 0$ takes place in cases when transformations $\mathrm{g}^{(1)}$ are the following: time reversal operation $1'$ and/or rotations around $k$-fold symmetry axes $k'_z$ ($k > 2$) and/or rotations around $n$-fold symmetry axes $n_z$ ($n > 1$) which exists simultaneously with reflection in plane $m_\perp$ orthogonal to the DW plane. Otherwise, the DWs with non-pyromagnetic point group have the magnetization distribution $\mathbf{M}(z) = -\mathbf{M}(-z)$. Only 64 from 125 magnetic point groups of DWs allow **M** [37]. They correspond to the magnetic DWs. As far as cubic $\mathrm{m\bar{3}m}$ crystal does not contain 6-fold symmetry axes (including inversion axes) it is necessary to exclude the magnetic point groups containing such symmetry elements. The remained set consists of the 57 magnetic point groups [38]. These groups are presented in table 1-3. The symbol (A,S) means that the specific function is the sum of odd and even function. The magnetic point groups which allow only two zero components of **M** or all odd components correspond to the cases of the $|\mathbf{M}| \neq const$. Such DWs appear near point of the phase transition [39].

The magnetic point group of the DW is the subgroup of the magnetic point group $G_P$ describing the symmetry of the crystal in the paramagnetic phase [39]. If influence of the crystal surfaces on the micromagnetic structure is taking into account then the group $G_P$ should be produced by the intersection of the magnetic point group $G_P^{\infty}$ (crystallographic class joined with transformation $1'$) and magnetic point group $G_S$ described the symmetry of the crystal surface. For the case of film or plate the $G_S$ is $\infty/\mathrm{mmm}1'$. The corresponding symmetry theory was described in [38].

Group-theoretical methods permit identification of the DW multiplicity [37]: the number of the DWs with identical energy and different structures. The DW multiplicity $q_k$ is determined by the relations of the group orders: $q_k = |G_P|/|G_k|$, were $G_k$ is the magnetic point group of the DW. If neighboring domains are determined then two different multiplicities appear: the DW multiplicity at the fixed boundary conditions $q_k' = |G_B|/|G_k|$ and the multiplicity of the boundary conditions $q_B = |G_P|/|G_B|$. Here the group $G_B$ is a magnetic point group of the boundary conditions (combination of two neighboring domains) [37,39]. The $G_B$ is the subgroup of the $G_P$ and is defined by type of DW and orientations of the domains magnetization directions in relation to the crystal surface [38]. It is worth to mention, that all possible groups $G_B$ are the same set as all possible groups $G_k$ (table 1-3). There are $q_k$ cosets and corresponding $q_k$ lost transformations $g_i^{(l)}$ (members of adjacent classes [37-40]) in series:

$$G_P = g_0^{(l)}G_k + g_1^{(l)}G_k + g_2^{(l)}G_k + \cdots + g_{q_k}^{(l)}G_k \,, \tag{1}$$

where $g_0^{(l)} \equiv 1$. These lost transformations are the transformations which relate all DWs with identical energies and different structures.

### 2.2. Application of the theory for description of the flexomagnetoelectric coupling

The local flexomagnetoelectric effect in the magnetic domain walls (DWs) of the cubic $\mathrm{m}\bar{3}\mathrm{m}$ crystal is described by the following expression [8,17,19,27]:

$$\mathbf{P} = \gamma_0[(\mathbf{M}\nabla)\mathbf{M} - \mathbf{M}(\nabla\mathbf{M}) + \cdots] \tag{2}$$

Here only the first in order of magnitude terms [12] are considered. Also, only short-range contribution [12] to the flexomagnetoelectric coupling was taken into account. In the case of the planar DW the expression (2) reduces to the following one-dimensional form:

$$\mathbf{P} = \gamma_0\left(M_z\frac{\partial\mathbf{M}}{\partial z} - \mathbf{M}\frac{\partial M_z}{\partial z}\right) \tag{3}$$

Thus, only transverse polarization components are nonzero:

$$P_x = \gamma_0 \left( M_z \frac{\partial M_x}{\partial z} - M_x \frac{\partial M_z}{\partial z} \right) \quad (4a)$$

$$P_y = \gamma_0 \left( M_z \frac{\partial M_y}{\partial z} - M_y \frac{\partial M_z}{\partial z} \right) \quad (4b)$$

Let us describe the relation between expression (3) and the group-theory predictions. If the function $M_i(z)$ is odd then the $\frac{\partial M_i(z)}{\partial z}$ is even, where $i = x, y, z$. If the function $M_i(z)$ is even then the $\frac{\partial M_i(z)}{\partial z}$ is odd. The product of two even or odd functions gives the even function. The product of even and odd function gives the odd function. These rules were used to obtain type of the electric polarization distribution produced by the flexomagnetoelectric coupling (2). The results do not correspond completely to the group-theory predictions. The $P_z(z)$ is always zero according to the (3) but the symmetry theory allows this component for all magnetic point groups (table 1-3). The same properties have also components $P_x(z)$ and $P_y(z)$ of the several specific DWs with the single component of the magnetization ($|\mathbf{M}| \neq const$): $k = 3, 4, 6, 44$ and 45 (table 3). Such components have been marked as 0(T), where "T" is the component type allowed by the symmetry. All remained magnetic point groups shows coincidence between the symmetry predictions and the expression (3) for the components $P_x(z)$ and $P_y(z)$.

The components of the electric polarization 0(T) allowed by the symmetry can be produced by the coupling type different from the known one (2): long-rage (non-local) contribution to the flexomagnetoelectric coupling (coupling via the spatially distributed stress tensor [1], demagnetization and depolarization energy terms, etc.), influence of the crystal surfaces (corresponding symmetry predictions were formulated in the [38]), considering of the next order in magnitude terms in the (2) [12], considering of the crystals with symmetry class non m$\bar{3}$m.

### 2.3. Chirality definition

The term "chirality" is widely used at the investigations of the magnetoelectric coupling [13,19] and properties of the spiral magnets [41,42]. Here according to the most popular definition, the chirality corresponds to the one of two possible opposite directions of the spin rotation axis. For example, chirality is defined [41] as the clockwise and counterclockwise rotation with respect to orientation of the Dzyaloshinsky-Moriya vector. However, it not always corresponds to the basic definition [43]. The ideal Néel DW and the spin cycloid structure are achiral according to the basic definition. The main reason of these confusions is ambiguity of this definition in magnetic symmetry [44]. The chirality-related problems cause appearance of the theory of the "complete symmetry" [45,46].

The introducing of the unified chirality definition to include motion was provided and time-invariant enantiomorphism has been defined [47-50]. It requires consideration of the enantiomeric states as interconverted by the space inversion, but not by the time reversal combined with any proper spatial rotation. If last part of requirement is violated then chirality is time-noninvariant. Object with this type chirality changes its enantiomeric states by not simultaneously applying of space inversion or time reversal transformation combined with any proper spatial rotation. In case of the time-invariant chirality the magnetic point group of enantiomorphs should not contain symmetry transformation $\overline{n}'$, where $n$ is the natural number or infinity [51]. Both time-invariant and time-noninvariant chiralities do not allow transformation $\overline{n}$ in the magnetic point group.

These criteria were applied for magnetic point groups in symmetry classification of the DWs. The magnetic point groups are divided into three tables (table 1-3) depend on their chirality.

### 3. Results and discussion

Different pairs of the groups $G_k$ and $G_B$ ($G_k \subseteq G_B$) correspond to the different types of the multiplicity of the DWs. If function $\mathbf{M}(z)$ is not the invariant of at least one lost symmetry transformation $g_i^{(l)}$ then multiplicity is ferromagnetic. If function $\mathbf{P}(z)$ is not the invariant of at least one lost symmetry transformation $g_i^{(l)}$ then multiplicity is ferroelectric. Otherwise, the multiplicity can be ferroelastic (the corresponding criteria are not considered here). There is a simple algorithm to determine such cases. If the magnetization components specified (table 1-3) for the $G_k$ and $G_B$ are different then multiplicity is ferromagnetic. The same rule exists for the polarization and ferroelectric multiplicity. Each of described multiplicity corresponds to the specific type of the Bloch lines [40].

There are two types of the structural transitions: with remaining of the magnetic point group and without it. Transition between DWs with the same symmetry and energy can be produced by the external influence that is not the invariant of all symmetry transformation $g_i^{(l)}$ which relates these two DWs. All possible changes of the DW symmetry correspond to the all subgroups $G_k$ of the group $G_B$:

$$1 \subseteq G_k \subseteq G_B \tag{5}$$

Thus, investigations of all group-subgroup relations in set $1 \leq k \leq 64$ (tables 1-3) leads to possibility to build complete map of structural (both magnetic and electric subsystem) transitions of magnetic DWs. The DWs with different magnetic point groups have different energies (such law is valid with high probability). Thus, transition with changing of the DW symmetry can be relaxation from metastable state to the state with lower energy.

Existing experimentally observation of the flexomagnetoelectric effect in the DW [13] agrees with presented symmetry theory. Corresponding experimental sample surface orientation is (210) and the DWs plane is (001). The magnetic point group of the crystal surface $G_S$ is $\infty_{[210]}/\mathrm{mmm}1'$. After its intersection with $G_P^{\infty} = \mathrm{m}\bar{3}\mathrm{m}1'$ we obtained $G_P = 2_{[001]}/\mathrm{m}1'$. The magnetic point group of the boundary conditions must be the subgroup of the $G_P$. The direction of the **M** in the domains are collinear with [210]. Then, according to the algorithm described in the [38] we have $G_B = 2'_z/\mathrm{m}_z$ ($k = 5$, table 3). The subgroups $G_k$ of this group describe symmetry of all possible DWs which can exist at such boundary conditions. They are the following: $2'_z$, $\bar{1}'$ and $\mathrm{m}_z$. Case $G_k = 2'_z$ describes the Bloch DW with asymmetrical magnetization distribution [39]. The DW with $G_k = \bar{1}'$ exists only near the point of phase transition. The DW with $G_k = \mathrm{m}_z$ ($k = 11$, table 3) has even component $M_z(z)$ and odd transverse components $M_x(z)$ and $M_y(z)$. Such symmetry requirements are satisfied by the DW observing in the [13]. The functions $P_x(z)$ and $P_y(z)$ of this DW are even that correspond to the observed uncompensated charges on the crystal surfaces. Corresponding dipole properties were detected as the response to the gradient of the electric filed. This DW has ferroelectric and ferromagnetic multiplicity at the same time. As well as this DW has only two states such property corresponds to the magnetoelectric properties of the DW. Here changing of the DW state corresponds to the changing of both ferroelectric and ferromagnetic structure. Two states of the DW structure which correspond to the two DW shift directions (effect signs [13]) are interrelated by the lost transformations $g_i^{(l)} = 2'_z$ or $\bar{1}'$:

$$G_B = G_k + \bar{1}' G_k = G_k + 2'_z G_k \tag{6}$$

The in-plane magnetic field as well as gradient of electric field is not the invariant of all these lost symmetry transformations. The hypothetic (allowed by the symmetry) odd component $\mathbf{P}_z(z)$ in this DW means presence of the coupled electrical charge in its volume. This component can exist only due to the mechanisms of the flexomagnetoelectric coupling different from the known one (2). These new mechanisms can be discovered by applying of external homogeneous electric field to the magnetic DWs in the cubic ferromagnetic crystal.

Also, new types of the flexomagnetoelectric coupling can be detected using the strong gradient of the electric filed applied to the achiral magnetic DWs with collinear magnetization which appear only near the points of phase transition: $k = 3, 4, 6, 44$ and 45 (table 3).

The whole discussed theory can be generalized to the weak ferromagnetics with non-collinear magnetic ordering as well as to the antiferromagnetics using algorithms specified in [52] and [39] respectively. Also, present symmetry classification predicts structural changes caused by the DW motion [39].

## 4. Conclusions

Thus, the magnetic point groups allow determining kind of the distributions of the electrical polarization in the domain wall volume produced by the flexomagnetoelectric effect. The time-noninvariant chiral domain walls have identical kind of spatial distribution of the magnetization and polarization. There are coincidence between the symmetry predictions and results obtaining from the known term of the flexomagnetoelectric coupling for transverse polarization components. Exceptions are only 5 magnetic point groups (total number is 57). They are achiral magnetic DWs with collinear magnetization which appear only near the points of phase transition. Applying of the homogeneous electrical field to the any magnetic DW in any magnetic material could lead to the observation of new flexomagnetoelectric properties.

**Acknowledgements**

I would like to express my sincere gratitude to my supervisor Prof. V.F. Kovalenko for him outstanding guidance, to Prof. V. A. L'vov and Dr. S. V. Olszewski for their helpful suggestions, to Dr. A. P. Pyatakov for productive discussion and to my wife D. M. Tanygina for spell checking.

**Table 1.** Types of spatial distribution of electric polarization induced by flexomagnetoelectric effect in magnetic DWs with the time-invariant chirality.

| $k$ | Magnetic point group | $M_x(z)$ | $M_y(z)$ | $M_z(z)$ | $P_x(z)$ | $P_y(z)$ | $P_z(z)$ |
|---|---|---|---|---|---|---|---|
| 7 | $2'_x 2_y 2'_z$ | A | S | 0 | 0 | 0 | 0(A) |
| 8 | $2'_z$ | A,S | A,S | 0 | 0 | 0 | 0(A,S) |
| 10 | $2'_x$ | A | S | S | S | A | 0(A) |
| 13 | $2_y$ | A | S | A | A | S | 0(A) |
| 16 | 1 | A,S | A,S | A,S | A,S | A,S | 0(A,S) |
| 19 | $2_z$ | 0 | 0 | A,S | 0 | 0 | 0(A,S) |
| 21 | $2_x 2_y 2_z$ | 0 | 0 | A | 0 | 0 | 0(A) |
| 24 | $3_z$ | 0 | 0 | A,S | 0 | 0 | 0(A,S) |
| 27 | $3_z 2_x$ | 0 | 0 | A | 0 | 0 | 0(A) |
| 30 | $4_z$ | 0 | 0 | A,S | 0 | 0 | 0(A,S) |
| 33 | $4_z 2_x 2_{xy}$ | 0 | 0 | A | 0 | 0 | 0(A) |
| 50 | $2_z 2'_x 2'_y$ | 0 | 0 | S | 0 | 0 | 0(A) |
| 53 | $3_z 2'_x$ | 0 | 0 | S | 0 | 0 | 0(A) |
| 57 | $4_z 2'_x 2'_y$ | 0 | 0 | S | 0 | 0 | 0(A) |

**Table 2.** Types of spatial distribution of electric polarization induced by flexomagnetoelectric effect in magnetic DWs with the time-noninvariant chirality:

| $k$ | Magnetic point group | $M_x(z)$ | $M_y(z)$ | $M_z(z)$ | $P_x(z)$ | $P_y(z)$ | $P_z(z)$ |
|---|---|---|---|---|---|---|---|
| 12 | $\mathrm{m}'_\mathrm{x}$ | 0 | A,S | A,S | 0 | A,S | 0(A,S) |
| 14 | $2_\mathrm{y}/\mathrm{m_y}'$ | A | 0 | A | A | 0 | 0(A) |
| 15 | $\bar{1}'$ | A | A | A | A | A | 0(A) |
| 17 | $\mathrm{m}'_\mathrm{x}\mathrm{m}'_\mathrm{z}2_\mathrm{y}$ | 0 | S | A | 0 | S | 0(A) |
| 18 | $\mathrm{m}'_\mathrm{z}$ | S | S | A | S | S | 0(A) |
| 20 | $2_\mathrm{z}/\mathrm{m}'_\mathrm{z}$ | 0 | 0 | A | 0 | 0 | 0(A) |
| 22 | $\mathrm{m}'_\mathrm{x}\mathrm{m}'_\mathrm{y}2_\mathrm{z}$ | 0 | 0 | A,S | 0 | 0 | 0(A,S) |
| 23 | $\mathrm{m}'_\mathrm{x}\mathrm{m}'_\mathrm{y}\mathrm{m}'_\mathrm{z}$ | 0 | 0 | A | 0 | 0 | 0(A) |
| 26 | $3_\mathrm{z}\mathrm{m}'_\mathrm{x}$ | 0 | 0 | A,S | 0 | 0 | 0(A,S) |
| 29 | $\bar{3}'_\mathrm{z}\mathrm{m}'_\mathrm{x}$ | 0 | 0 | A | 0 | 0 | 0(A) |
| 31 | $4_\mathrm{z}/\mathrm{m}'_\mathrm{z}$ | 0 | 0 | A | 0 | 0 | 0(A) |
| 32 | $4_\mathrm{z}\mathrm{m}'_\mathrm{x}\mathrm{m}'_\mathrm{xy}$ | 0 | 0 | A,S | 0 | 0 | 0(A,S) |
| 34 | $4_\mathrm{z}/\mathrm{m}'_\mathrm{z}\mathrm{m}'_\mathrm{x}\mathrm{m}'_\mathrm{xy}$ | 0 | 0 | A | 0 | 0 | 0(A) |
| 35 | $\bar{4}'_\mathrm{z}$ | 0 | 0 | A | 0 | 0 | 0(A) |
| 36 | $\bar{4}'_\mathrm{z}2_\mathrm{x}\mathrm{m}'_\mathrm{xy}$ | 0 | 0 | A | 0 | 0 | 0(A) |
| 42 | $\bar{3}'_\mathrm{z}$ | 0 | 0 | A | 0 | 0 | 0(A) |

**Table 3.** Types of spatial distribution of electric polarization induced by flexomagnetoelectric effect in magnetic achiral DWs.

| $k$ | Magnetic point group | $M_x(z)$ | $M_y(z)$ | $M_z(z)$ | $P_x(z)$ | $P_y(z)$ | $P_z(z)$ |
|---|---|---|---|---|---|---|---|
| 1 | $m_x m_z m'_y$ | A | 0 | 0 | 0 | 0 | 0(A) |
| 2 | $m'_y m_x 2'_z$ | A,S | 0 | 0 | 0 | 0 | 0(A,S) |
| 3 | $m_x m_z 2_y$ | A | 0 | 0 | 0 | 0(S) | 0(A) |
| 4 | $2'_x/m_x$ | A | 0 | 0 | 0 | 0(A) | 0(A) |
| 5 | $2'_z/m_z$ | A | A | 0 | 0 | 0 | 0(A) |
| 6 | $m_y$ | A,S | 0 | 0 | 0 | 0(A,S) | 0(A,S) |
| 9 | $m_z m'_y 2'_x$ | A | 0 | S | S | 0 | 0(A) |
| 11 | $m_z$ | A | A | S | S | S | 0(A) |
| 43 | $m_y m'_x m'_z$ | 0 | S | 0 | 0 | 0 | 0(A) |
| 44 | $m_y m'_z 2'_x$ | 0 | S | 0 | 0(S) | 0 | 0(A) |
| 45 | $2_y/m_y$ | 0 | S | 0 | 0(A) | 0 | 0(A) |
| 46 | $2'_z/m'_z$ | S | S | 0 | 0 | 0 | 0(A) |
| 47 | $2'_x/m'_x$ | 0 | S | S | 0 | A | 0(A) |
| 48 | $\bar{1}$ | S | S | S | A | A | 0(A) |
| 49 | $2_z/m_z$ | 0 | 0 | S | 0 | 0 | 0(A) |
| 51 | $m_z m'_x m'_y$ | 0 | 0 | S | 0 | 0 | 0(A) |
| 55 | $\bar{3}_z m'_x$ | 0 | 0 | S | 0 | 0 | 0(A) |
| 56 | $4_z/m_z$ | 0 | 0 | S | 0 | 0 | 0(A) |
| 58 | $4_z/m_z m'_x m'_{xy}$ | 0 | 0 | S | 0 | 0 | 0(A) |
| 59 | $\bar{4}_z$ | 0 | 0 | S | 0 | 0 | 0(A) |
| 60 | $\bar{4}_z 2'_x m_{xy}{}'$ | 0 | 0 | S | 0 | 0 | 0(A) |
| 64 | $\bar{3}_z$ | 0 | 0 | S | 0 | 0 | 0(A) |